*Communication*

# Direct In-Situ Capture, Separation and Visualization of Biological Particles with Fluid-Screen in the Context of Venus Life Finder Mission Concept Study

Robert E. Weber [1,2], Janusz J. Petkowski [3] and Monika U. Weber [1,*]

[1] Fluid-Screen, Inc. 100 Cummings Center, Suite 243-C, Beverly, MA 01915, USA
[2] Hener, Wrocław Technology Park, BETA Building, Room 104, Klecińska 125, 54-413 Wrocław, Poland
[3] JJ Scientific, 02-792 Warsaw, Poland
[*] Correspondence: monika.weber@gmail.com

**Abstract:** Evidence of chemical disequilibria and other anomalous observations in the Venusian atmosphere motivate the search for life within the planet's temperate clouds. To find signs of a Venusian aerial biosphere, a dedicated astrobiological space mission is required. Venus Life Finder (VLF) missions encompass unique mission concepts with specialized instruments to search for habitability indicators, biosignatures and even life itself. A key in the search for life is direct capture, concentration and visualization of particles of biological potential. Here, we present a short overview of Fluid-Screen (FS) technology, a recent advancement in the dielectrophoretic (DEP) microbial particle capture, concentration and separation. Fluid-Screen is capable of capturing and separating biochemically diverse particles, including multicellular molds, eukaryotic cells, different species of bacteria and even viruses, based on particle dielectric properties. In this short communication, we discuss the possible implementation of Fluid-Screen in the context of the Venus Life Finder (VLF) missions, emphasizing the unique science output of the Fluid-Screen instrument. FS can be coupled with other highly sophisticated instruments such as an autofluorescence microscope or a laser desorption mass spectrometer (LDMS). We discuss possible configurations of Fluid-Screen that upon modification and testing, could be adapted for Venus. We discuss the unique science output of the Fluid-Screen technology that can capture biological particles in their native state and hold them in the focal plane of the microscope for the direct imaging of the captured material. We discuss the challenges for the proposed method posed by the concentrated sulfuric acid environment of Venus' clouds. While Venus' clouds are a particularly challenging environment, other bodies of the solar system, e.g., with liquid water present, might be especially suitable for Fluid-Screen application.

## 1. Introduction

Dielectrophoresis (DEP) is the unidirectional motion of neutral, polarizable particles suspended in a fluid triggered by the presence of a non-uniform electric field (Figure 1).

DEP has been proposed as a biological particle capture and separation method based on the particles' dielectric properties [1–3]. The use of DEP technology to manipulate particle and cell motion has experienced great advancements in terms of versatility and flexibility and has been explored for decades (see e.g., [1,4–11], more recently reviewed in [5,6,12]). The modern utilization of the DEP phenomenon focuses on the isolation, manipulation, and concentration of bioparticles using dielectrophoretic force in miniaturized microfluidic devices [6,13,14].

Fluid-Screen (FS) is a recent advancement in the DEP-based biological particle capture and separation technology (Figure 2) [15–17]. Fluid-Screen is capable of capturing and separating very diverse biological particles, including multicellular molds, eukaryotic cells, different species of bacteria and even viruses [17] (Figure 3a). Fluid-Screen DEP technology can also capture and separate cells at different stages of their respective life cycles (Figure 3b). While the capture and separation of large biomolecules, such as DNA or proteins, is not a direct focus of the FS technology it can in principle be adapted to capture and separate them as well [18–22].

Here, we present a short perspective article on the possible future application and adaptation of Fluid-Screen DEP technology to the astrobiological exploration of Venus and other bodies in the Solar System. We start by providing an overview of Fluid-Screen technology (Section 1) followed by the motivation for the use of the FS System for direct capture and imaging of particles of possible biological origin in Venus clouds. We describe the proposed concept for Fluid-Screen for Venus Life Finder mission concept study (Section 3) and discuss the challenges for the adaptation of dielectrophoretic biological particle capture in the unique Venusian environment, focusing in particular on the challenges posed by the concentrated sulfuric acid (Section 4).

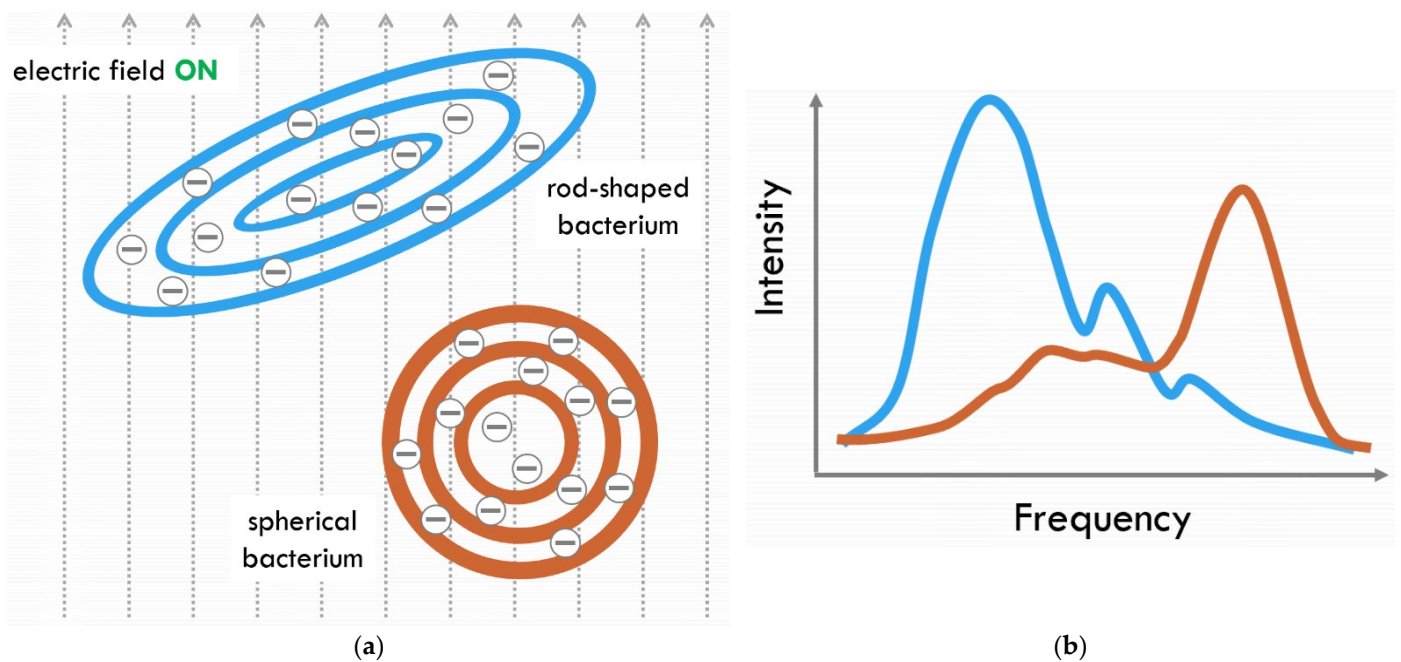

**Figure 1.** The schematic representation of the principle behind Fluid-Screen DEP technique. (**a**) Electric field induces a dipole in every building block of a bacterium (we call a sum of all of these dipoles "super dipole"). The super dipole depends on shape, size, morphology and chemistry of the particle; Biological particles suspended in the liquid sample are captured and concentrated on the basis of super dipole interacting with an electric field. Such interaction leads to very fast, efficient, and reliable capture and separation of a variety of microbial cells from other particles (particle size is not a key factor, thus FS can concentrate across sizes). (**b**) Fluid-Screen DEP response spectrum (schematic represention) is different for each bacterial species due to differences in shape, size, morphology and chemistry of the particle; this effect allows for selective capture and separation of microorganisms.

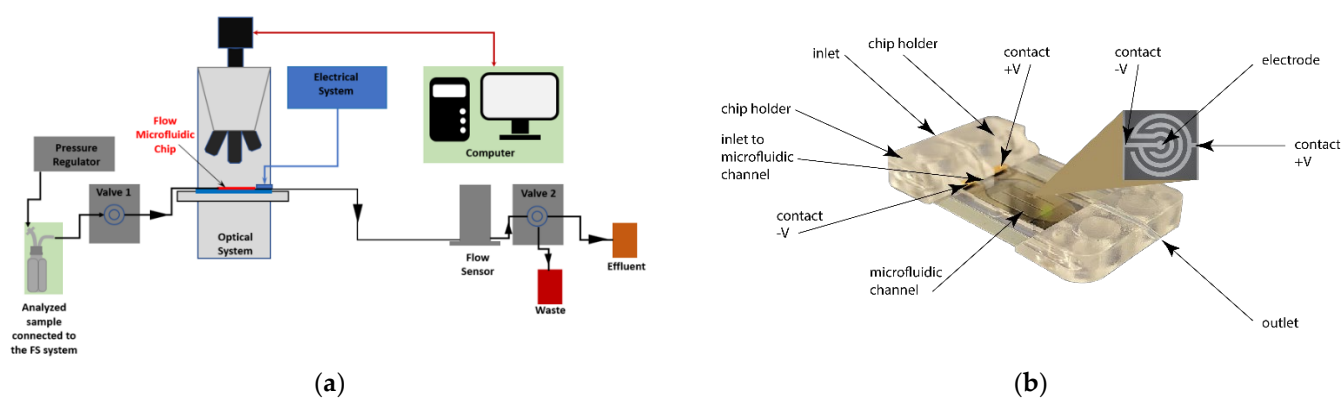

(**a**) (**b**)

**Figure 2.** The schematic of the Fluid-Screen (FS) microbial capture and separation system. (**a**) The overall schematic of the FS System operation. The influent sample enters the chip. When the electric field is turned on, bacteria are captured on the electrodes (Figure 4). After the electric field is turned off, the effluent sample is collected. Sample processing of 1 mL through FS System takes approx. 4 min. In addition, Fluid-Screen has developed an automated method to detect and quantify the number of captured biological particles. Figure 2a from [17], reproduced under a Creative Commons Attribution 4.0 International License. (**b**) The Schematic of the FS Flow Microfluidic Chip with a visible electrode.

*Fluid-Screen Technology Overview*

The Fluid-Screen technology controls particle motion by dielectrophoresis (DEP). This universal phenomenon describes the motion of all particles in a non-uniform electric field gradient. Fluid-Screen has successfully used this technique to capture and concentrate particles, bacteria and other cells [17]. Universal capture allows capturing and concentrating all particles within a range of sizes, including viruses, from diverse sample matrices.

The key innovation of Fluid-Screen is a unique ring-shaped electrode design to maximize the bacterial response to the electric field (Figure 2) [15–17]. Fluid-Screen uses dielectrophoresis force to capture particles on electrodes inside the chip and subsequently image captured biological particles on the electrodes with a camera mounted to the microscope (Figure 2). There are two advantages to this approach.

Biological particles captured on the electrodes are immobilized and do not move in $x$, $y$, and $z$ directions, i.e., particles do not show Brownian motion. This enables imaging and detecting particles. In the absence of capture, particles are free to move in the $x$, $y$, and $z$ directions. Movement in the $z$ direction causes particles to go out of focus, which means particles are not detected. Figure 4 below illustrates the capture and visualization of *Aspergillus brasiliensis* cells on the FS ring-shaped electrode system. Particles about 4 μm in size are captured in their native state, in bright field (no fluorescence, no staining). The same sample is imaged with an electric field off (no capture, free particle movement in $x$, $y$, and $z$ directions) and with a field on (capture, particles are held in place, in the focal plane of the microscope). In the absence of capture, such as with a conventional microscope, particles may be present in the sample, but they are not detected, because the particles are not in focus.

Fluid-Screen achieves selectivity of capture due to the differences in the dielectric response of particles. The dielectric response depends on the chemical material the particle is made of and particle morphology. Size is not the primary factor in selectivity.

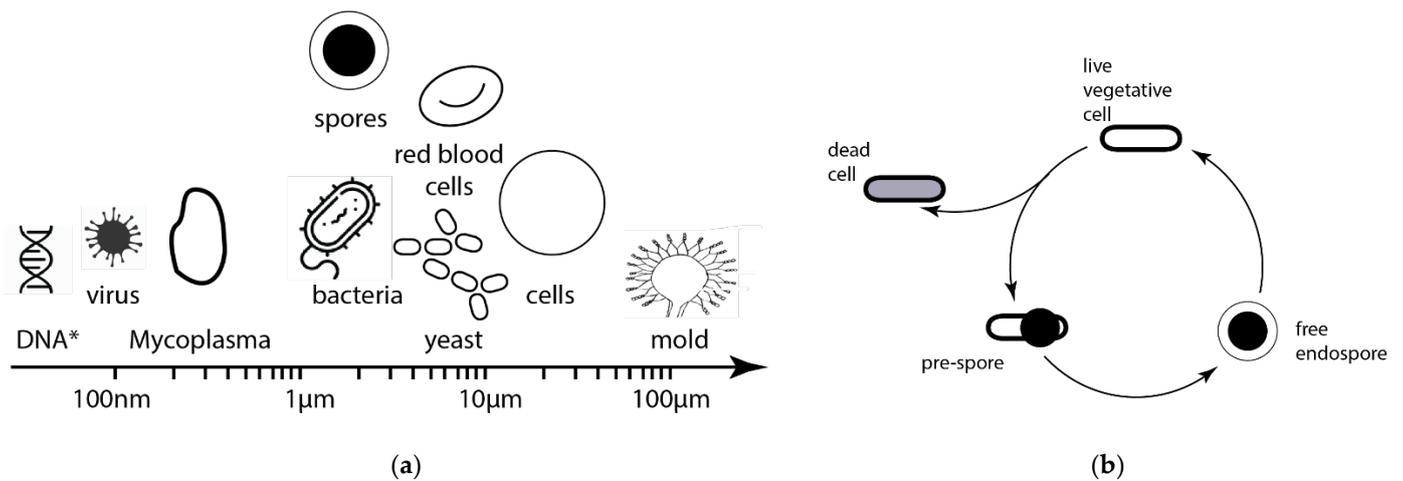

(**a**) (**b**)

**Figure 3.** The Fluid-Screen technology is universal and can capture and separate different species of bacteria, fungi, and viruses from various sample matrices. (**a**) FS technology captures and separates diverse life forms. * Capturing of DNA and other large biomolecules has been demonstrated in the literature [21,22] and it is not a direct focus of Fluid-Screen, although FS technology could be adapted to capture large biomolecules as well. (**b**) FS technology can also differentiate between different physiological stages of life and can selectively capture life forms at their different physiological stages.

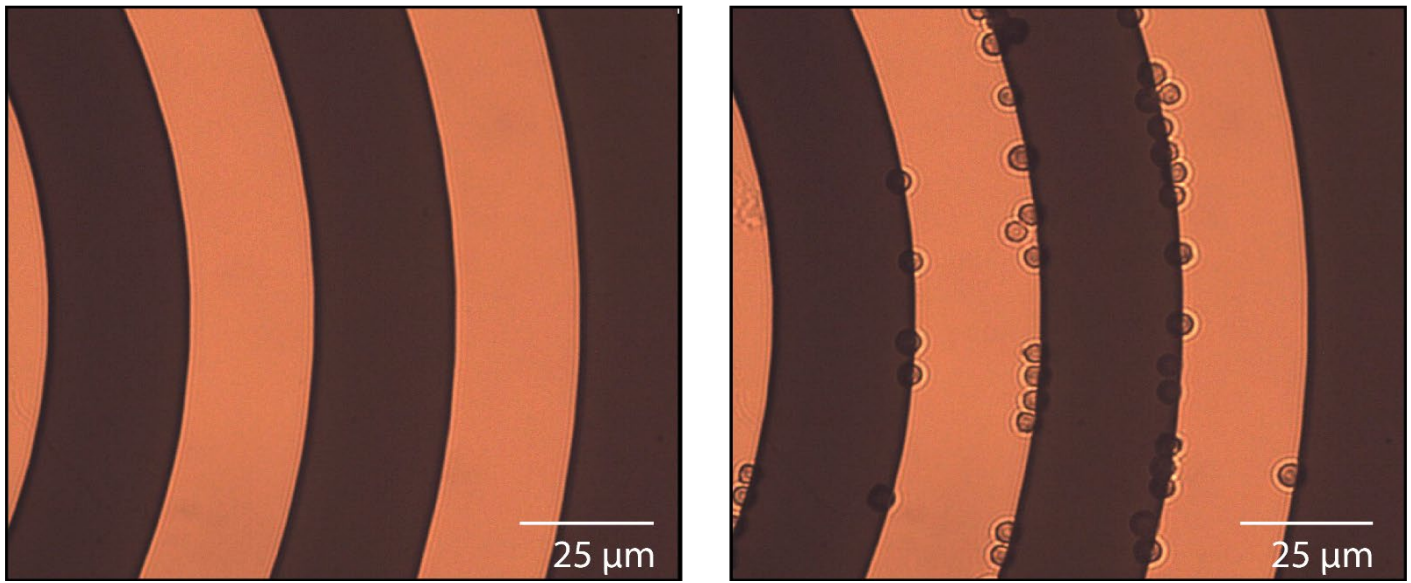

**Figure 4.** Bright field images show a section of the Fluid-Screen Chip's circular ring-shaped electrode structure. Electric field off (left panel), electric field on (right panel). The cells are captured from deionized water spiked with *Aspergillus brasiliensis* suspension. When the electric field is off (left panel), there is no bacterial capture observed and *Aspergillus brasiliensis* (large size dark spheres) are suspended in the solution, which appears out of focus. After the electric field is applied, (right panel), *Aspergillus brasiliensis* cells come into focus and are visibly captured on the electrode edges and are held in place, in the focal plane of the microscope.

## 2. Motivation for Utilization of Fluid-Screen System in the Exploration of Venus' Clouds

The ability of the FS System to capture, concentrate and separate a variety of different biological particles has tremendous applications in many branches of human industry. Space exploration and space missions where science objectives require biological particle capture, sorting, concentration, detection and imaging are no different. FS shows promise especially in astrobiological space missions where direct life detection and imaging are required. If life is present elsewhere in the solar system it is quite likely that such life particles will be exceedingly rare and difficult to capture. Sorting through a very large number of non-biological particles just to find a dozen or so candidate cell-like structures that could be targeted for further chemical analysis or imaging is a great challenge and requires novel technological inventions and approaches that have not been introduced to the space sector before. Fluid-Screen is the first universal biological particle capture, concentration and separation system that is considered for Venus, or any other astrobiology-focused mission.

Life beyond Earth might have significantly different biochemistry including different metabolism, genetics, cell walls or membranes. As long as biological particles respond to the electric field FS can find suitable conditions to capture them. FS can capture a diverse collection of cells, and other biological particles, sort them (i.e., selectively capture particles with certain pre-determined properties) and eventually concentrate the rare sub-population of particles of biological interest, enabling their detail testing or visualization by either a dedicated microscope or a coupled mass-spectrometer for chemical analysis of captured particles. FS technology is not limited by the scarcity of potential biological particles, it can capture single cells of a specific species even from very diluted samples, or complex mixtures containing many different cells and particles [17]. This advantage of FS is especially important for Venus clouds. Venus clouds are the only potentially habitable environment on the planet, with the surface and the subsurface being too hot for any kind of complex organic chemistry that could build life. This limitation in available habitat means that if the clouds of Venus are inhabited by any life, it is likely very scarce, with a biomass density much lower (possibly by orders of magnitude lower) than the biomass found in the harshest of the inhabited environments on Earth. Therefore, by definition, the search for life in the clouds of Venus is like a search for a proverbial "needle in a haystack". As shown below, FS is uniquely suited for such a challenging task (Section 3).

For the Venus Life Finder (VLF) [23], we envision an FS Chip configuration, a complex FS System with microfluidics and an autofluorescence microscope (Section 3) that is suitable for the VLF missions that have a larger payload allowance (e.g., Venus Airborne Investigation of Habitability and Life (VAIHL) mission described in [23], as well as in companion papers in the same issue [24–27]). We describe the FS Chip configuration below.

## 3. Fluid-Screen Chip Configurations for Venus Life Finder VAIHL Mission

If the measurements of the Pioneer Venus are correct, then there are three distinct populations of particles, called "Modes", that form the clouds of Venus (Table 1) [28–33]. The first population is small Mode 1 particles with a mean diameter of 0.4 μm. Mode 2 particles, with a diameter of ~2 μm, are spherical liquid droplets of concentrated sulfuric acid. The large, ~8 μm [29,30], possibly non-spherical particles of unknown composition that could be a solid or semi-solid salt slurry form the Mode 3 population [34].

**Table 1.** Distribution of the Venusian cloud particles. Data originally from [32,33], based on review of the data in [23]. (1)—Mode 1 particles; (2)—Mode 2 particles; (3)—Mode 3 particles.

| Region | Altitude (km) | Temperature (K) | Pressure (atm) | Cloud Particle Properties | | |
|---|---|---|---|---|---|---|
| | | | | *Average Num. Density* (n cm$^{-3}$) | *Mean Diameter* (μm) | *Cloud Particle $H_2SO_4$ Concentration* |
| Layers above upper haze | 100–110 | | | | | 100% $H_2SO_4$ |
| Upper haze | 70–90 | 225–190 | 0.04–0.0004 | 500 | 0.4 | 70% $H_2SO_4$ 30% $H_2O$ |
| Upper cloud | 56.5–70 | 286–225 | 0.5–0.04 | (1) −1500 (2) −50 | Bimodal 0.4 and 2.0 | liquid 80% $H_2SO_4$ 20% $H_2O$ |
| Middle cloud | 50.5–56.5 | 345–286 | 1.0–0.5 | (1) −300 (2) −50 (3) −10 | Trimodal 0.3, 2.5 and 7.0 | liquid 90% $H_2SO_4$ 10% $H_2O$ |
| Lower cloud | 47.5–50.5 | 367–345 | 1.5–1.0 | (1) −1200 (2) −50 (3) −50 | Trimodal 0.4, 2.0 and 8.0 | liquid 98% $H_2SO_4$ 2% $H_2O$ (or fumic $H_2SO_4$) |
| Lower haze | 31–47.5 | 482–367 | 9.5–1.5 | 2–20 | 0.2 | |
| Pre-cloud layers | 46 and 47.5 | 378 and 367 | 1.8–1.5 | 50 and 150 | Bimodal 0.3 and 2.0 | |

**Table 2.** An overview of the FS Chip configurations of Fluid-Screen with potential application to VLF missions.

| Instrument Characteristics | Fluid-Screen Chip |
|---|---|
| Mass (kg) | 1–10 |
| Volume (cm$^3$) | 3000 |
| Power (W) (average/peak) | 1.5–30 |
| Data volume (per measurement and total) | 100 kb–2.4 mb per image, max 20 images per second * |
| Data sampling time | <1 s per image * |

* Depends on the microscope and the camera model.

The FS Chip configuration aims to capture and image any biological particles suspended in the Venus cloud droplets (Figure 5). The particles will be either directly observed (for particle size >0.5 μm) or observed as aggregates (for particle size <0.5 μm).

*3.1. Fluid-Screen Chip Characteristics*

The Fluid-Screen Chip is a design concept for biological particle capture from the atmosphere. In this configuration, the FS System is envisioned to be an integral part of the VAIHL mission concept [23] and would work together with the autofluorescence microscope [35,36] and/or together with LDMS (e.g., [37]) (Figure 5).

The FS Chip configuration provides versatility with connections to the other science instruments but requires a separate sample collector that collects the liquid sulfuric acid droplets to the primary reservoir. The inlet tube then needs to connect the FS Chip to the primary sample reservoir. Such sample collector technology is a challenge for Venus but could, in principle, derive from cloud droplet technologies developed for Earth, e.g., using the modified "fog harp" technique (e.g., [38–41]). Collecting samples in bulk could also perturb the unique conditions of the microenvironment of the individual droplets and hence destroy any potential biological particles before FS capture could commence (see Section 4 for the discussion of challenges).

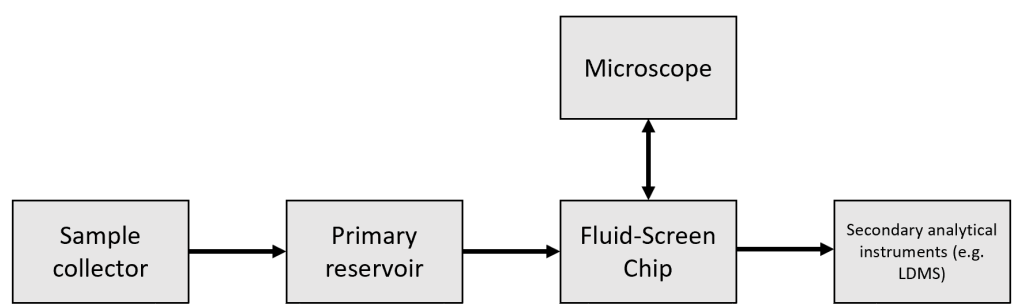

**Figure 5.** Schematic of the Fluid-Screen in the context of other instruments on the VAIHL mission concept. Liquid cloud droplets are collected by sample collector (e.g., analogous to the "fog harp" instrument [38–41]) into the primary reservoir which then feeds the sample to FS Chip. FS Chip is combined with a microscope for direct visualization of captured biological particles. Upon successful visualization of the sample Fluid-Screen can release the captured particles for analysis in secondary instruments (e.g., LDMS [37]).

FS System output will be concentrated samples delivered to a mass spectrometer for analysis (e.g., [37]) or directly imaged by a separate microscope system (e.g., [35,36]). The combination of FS with the microscope system would allow for directly detecting and characterizing the morphological indicators of life, which is one of the main science objectives of the VLF VAIHL mission [23].

Any direct detection of life's structures (e.g., cell-like vesicles and other life particles) requires an imaging system. A microscope is required to directly detect and characterize morphological indicators of life. Any kind of life, no matter its chemical makeup, likely needs some barriers, e.g., cell walls, cell membranes, etc., that allow it to exist in a distinct form separated from the surrounding environment (i.e., allow it to maintain a cell-like structure). FS System collects, captures, sorts and concentrates any biological particles that might reside in the collected liquid cloud particles and holds them at the focal plane of the microscope to properly image the sample (Figure 2). The FS System is developed to work with virtually any optical microscope system.

One choice microscope for coupling with Fluid-Screen is a UV-autofluorescence microscope. On Earth, many natural compounds are known to fluoresce when subjected to UV radiation. Autofluorescence allows for the detection of organics. Detecting autofluorescence is in fact one of the main science objectives of the Autofluorescence Nephelometer (AFN) instrument [42] selected for the upcoming Rocket Lab mission to Venus [43]. Autofluorescence could also be utilized for the detection of organic molecules within the captured cloud particles and enable the direct detection and label-free imaging of microbes and other cell-like structures that contain fluorescing organic material. An example of an autofluorescence microscope system that could be adapted for such a task is SHERLOC (which is a part of the Perseverance Rover currently at Mars) [35,36]. SHERLOC includes a UV-autofluorescence Raman spectroscopy system that operates at 30 μm resolution in a wide-field scanning capacity. However, in an alternative optical configuration, deep UV fluorescence microscopy can achieve resolutions in the sub-μm resolution and would detect native UV fluorescence of any organics (<1 ppb of aromatic organics) within captured cloud particles [35,36]. The detection of organics with the Fluid-Screen and UV-autofluorescence system does not require any externally added reagents or fluorescent dyes and solely exploits UV (<250 nm) excitation and 320 nm emission.

FS can also be coupled with the LDMS. FS captures particles and concentrates them, once captured the particles can be released from the FS System and presented to the LDMS. FS can capture particles of selected DEP properties which allows for the selective chemical analysis of well-defined sub-populations of particles. As with any sample capture approaches for LDMS [37] more work is needed to develop an interface connecting the microfluidics of the FS to the LDMS interface (Section 4).

*3.2. Predicted Scientific Outcome for Venus VAIHL Mission*

The scientific outcome of including the Fluid-Screen System in the Venus mission is the selective capturing and concentrating of any biological particles present in the liquid phase of the clouds and providing images of particles captured directly from Venus Clouds (with the aid of a dedicated microscope).

We estimate the number of particles the FS System can capture per hour of operation under the following assumptions. We estimated the number of particles that would be captured on FS Chip following the results of Pioneer Venus from (Table 1) [32,33] and assumptions of the flow rate and sampling time from the Venus Life Finder Mission Study [23]. In addition, under low power requirements (1.5 W–30 W) and a low flow rate of 0.1 mL/min FS can capture up to 99% of particles in test samples, using standard FS capture protocols [17]. The efficiency of capture of particles from concentrated sulfuric acid remains to be assessed and is likely much lower and requires different and uniquely tailored capture protocols (Section 4); even if the capture efficiency is a fraction of standard test conditions (e.g., 1%) it still would result in the capture of hundreds of biological particles, especially for longer duration missions (Figure 6).

| Assumptions | Low Power (5W-30W) | Flow Chip | | Low Power (5W-30W) | Flow Chip | |
|---|---|---|---|---|---|---|
| | Flow rate [mL/min] | | 0.1 | Flow rate [mL/min] | | 0.1 |
| | Chip capture efficiency | | 99% | Chip capture efficiency | | 99% |
| Mission duration in days | 1 | 41 | 365 | 1 | 41 | 365 |
| Tested volume [mL] | 144 | 5907 | 52,560 | 1440 | 59,040 | 525,600 |
| Mode 1 particles collected | 1426 | 58,480 | 520,344 | 14,256 | 584,496 | 5,203,440 |
| Mode 1 biological particles collected | 3 | 117 | 1041 | 29 | 1169 | 10,407 |
| Mode 2 particles collected | 5702 | 233,918 | 2,081,376 | 57,024 | 2,337,984 | 20,813,760 |
| Mode 2 biological particles collected | 342 | 14,035 | 124,883 | 3421 | 140,279 | 1,248,826 |
| Mode 3 particles collected | 4063 | 166,667 | 1,482,980 | 40,630 | 1,665,814 | 14,829,804 |
| Mode 3 biological particles collected | 244 | 10,000 | 88,979 | 2438 | 99,949 | 889,788 |

**Figure 6.** The Fluid-Screen capture efficiency depends on the power requirements and the flow rate [17]. For example, under low power requirements (1.5 W–30 W) and high flow rate of 1 mL/min the capture efficiency in standard FS conditions [17] is only 30%. The capture efficiency raises dramatically if the flow rate is slow, 0.1 mL/min, or more power is provided to the system. The calculated capture efficiency does not consider the potential challenges posed by the sulfuric acid environment of the clouds (Section 4). Such challenges to the performance of FS System would have to be assessed separately by a dedicated experimental laboratory program. Estimated number of biological particles that would be captured on Fluid-Screen Chip follows the assumptions of the *Venus Sampling—Flow Rate and Sampling Time Calculation* model developed by prof. Christopher E. Carr and Dr. Rachel Moore for the VLF Mission Study [23].

Note, that the numbers presented in Figure 6 are notional and are likely to change, as the mission concepts keep getting developed, depending on the sample collector technology implemented, collection efficiency or the underlying assumptions of the model. For example, a less efficient 30 × 30 cm passive sample collector would be able to acquire approx. 10 milliliters of liquid in lower clouds per day of sampling and less than 0.5 milliliters per day in upper cloud layers. Under such assumptions, one milliliter of collected liquid contains $10^9$–$10^{13}$ droplets [27].

**4. Challenges and Path Forward**

There are several significant challenges that need to be investigated before the adaptation of the Fluid-Screen System to the Venusian cloud environment. We discuss them below.

*4.1. Dielectrophoretic Capture of Particles in Concentrated Sulfuric Acid*

Particle capture via dielectrophoresis (DEP) happens when a particle has a differential polarization with respect to the surrounding environment (i.e., the solution in which it is suspended) (Figure 7). Moreover, the part of the dielectrophoresis equation (Figure 7) related to the conductivity of the particle and the medium reaches the maximum at $\sigma_m/\sigma_p$ = 0.4, where $\sigma_m$ is the conductivity of the medium and $\sigma_p$ is the conductivity of the particle. In theory, this effect is independent of the medium used [44] and can be applied to a variety of liquids including concentrated sulfuric acid.

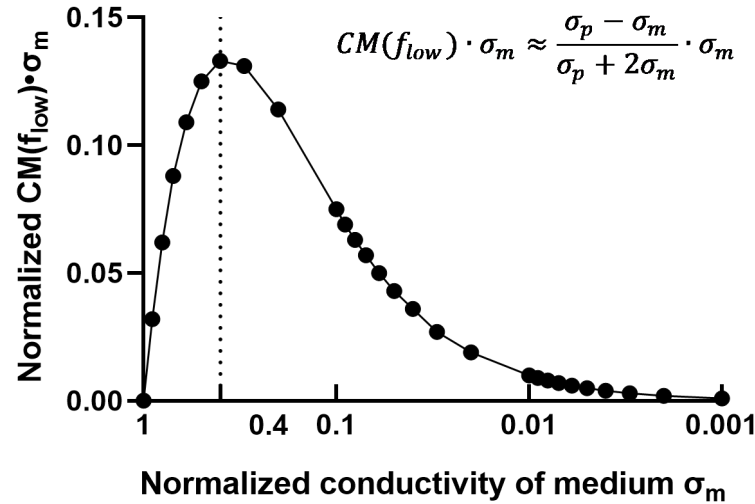

**Figure 7.** Normalized Clausius–Mossotti (CM) factor multiplied by conductivity of the medium versus normalized conductivity of medium in the low frequency range below 100 MHz. The conductive behavior dominates the value of the CM factor and the DEP force is attractive (see [44], their Section 2.4.1).

The challenge for the FS System is to sustain enough electric field in the presence of a high conductivity medium (such as concentrated sulfuric acid) which strengthens the effects of the short (leakage current) between the electrodes. The challenge for DEP in high-conductivity media is not an absolute limitation that stems from the underlying physics of the DEP process but rather a technological one. The conductivity of 70% $H_2SO_4$ is 231.5 mS/cm which is 15 times higher than undiluted phosphate-buffered saline PBS, the optimal environment for microorganisms on Earth (Figures 7 and 8). The preliminary calculation presented in Figure 7 suggests that the dielectrophoretic capture in high conductivity solutions could be attainted upon the modification of the FS System and is the focus of future work. The immediate solution to the challenge of high-conductivity solutions would be to dilute the concentrated sulfuric acid medium to reach the optimal value of 0.4 (Figure 7) this, however, increases the volume of the sample and changes the native chemical environment of any particles suspended in the liquid.

Experimental tests of DEP with the FS System have to be performed to fully understand the limitations posed by this aggressive solvent. Sample testing and modeling in various concentrations of sulfuric acid for compatibility with the chip and to confirm the operating principle of dielectrophoresis would need to be performed.

For example, it might be the case that the hypothetical "cells" of Venusian microbes have lower conductivity in the cell interior compared to the exterior concentrated sulfuric acid environment (due to e.g., active mechanisms of water accumulation and retention). If this is the case then the opposite effects are expected to happen and instead of DEP attraction to the electrode and particle capture, the repellence of particles away from the electrode occurs. If such a phenomenon happens, a significant modification of the architecture of the FS System presented here would be required.

This problem is also related to the challenge of the unambiguous identification of the captured material and the possibility of false positives. The exact chemical composition of cloud particles is unknown which might pose significant difficulties in the interpretation of the obtained data, ruling out possible false positives and distinguishing them from the true microbial-type life. This problem, however, is not unique to Venus and is universal to all astrobiological exploration of the Solar System. Specifically, cell-like structures or other morphological signs of life cannot be used as evidence of life on their own, as under the microscope many abiotic formations tend to look very similar to cell-like structures [45]. As mentioned above, the detection of any morphological signs of life has to be followed up by a detailed chemical analysis of the candidate structures. Nevertheless, Fluid-Screen can help in distinguishing between real cells and their abiotic mimics. For example, if a particle is a microbial cell coated with a cyclooctasulfur ($S_8$) cell wall, as it was previously hypothesized for the Venusian aerial biosphere [46], it would experience negative or positive dielectrophoresis, depending on the chemical composition of the interior of such cell. On the other hand, an abiotic mimic of such sulfur-coated cells, i.e., a solid sulfur particle that would probably look identical under the microscope, would experience no DEP force or a very weak DEP force. Such differences in particle behavior could help in the interpretation of the results and further minimize the risk of false-positive detections.

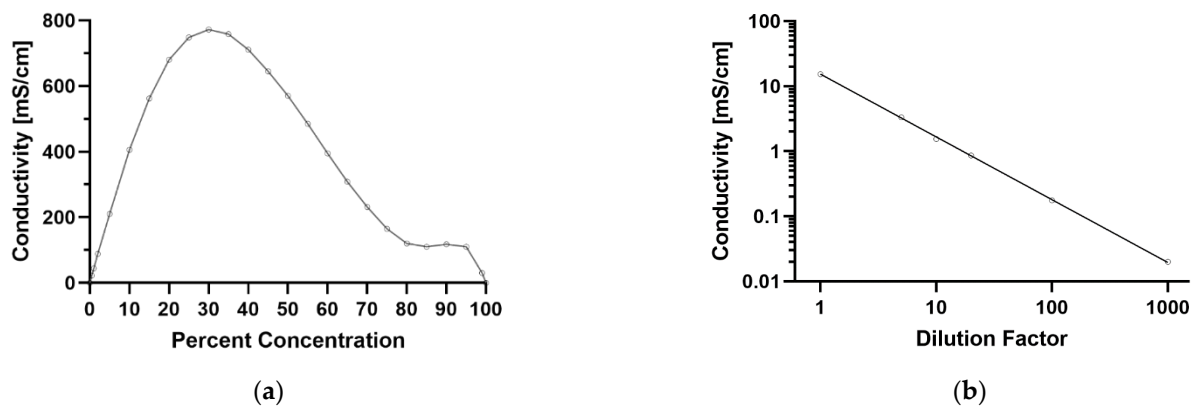

**Figure 8.** (**a**) Conductivity of concentrated sulfuric acid solution at 21 °C, at different concentrations. The conductivity of the liquid concentrated sulfuric acid (70–100%) in the clouds of Venus is in the range of 100 mS/cm, much higher than conductivities usually used in the DEP microbial capture methods. Such high conductivity poses a challenge for the DEP capture technique. Conductivity values for concentrated sulfuric acid solutions calculated based on values from [47,48]. (**b**) Phosphate buffered saline (PBS) at different dilution factors, diluted with MiliQ water at 21 °C. Diluted PBS (20 µS/cm) is the standard buffer used in FS DEP microbial capture, concentration and separation [17].

The exact chemical composition of Venus' cloud droplets is unknown. To work FS System needs a liquid sample (to capture any potentially suspended biological solid particles in the solution). The clouds of Venus likely are heterogeneous in nature and could be a mixture of various populations of particles, including liquid droplets of concentrated sulfuric acid (i.e., Mode 2) and solid or semi-solid particles of irregular shape (i.e., Mode 3) [34]. The capture of pure solids devoid of the liquid medium is not possible. Clogging of the FS microfluidics system with solid particulate matter might also be a challenge that would require the primary collector to be designed to capture exclusively liquid droplets or solid particles suspended in a liquid.

In contrast to the experimental challenges to the DEP biological particle capture posed by the concentrated sulfuric acid conditions, the adaptation of the FS System engineering itself to the Venusian cloud conditions is much more straightforward. The core of the FS System is a chip made of glass and gold, which is compatible with sulfuric acid. The remaining parts of the system would need to be protected, and utilization of such protective material including Teflon is feasible. To adapt to operation in sulfuric acid, material compatibility tests with sulfuric acid need to be performed (this includes pumps, tubing, and other subsystems). On the basis of the compatibility tests with concentrated sulfuric acid new electronic capture protocols would need to be developed.

We note that while Venus' sulfuric acid clouds are a particularly challenging environment for DEP-based biological particle capture methods, other bodies of the solar system with liquid water present, such as Jupiter's moon Europa or Saturn's moon Enceladus is much less challenging and will be a subject of a separate dedicated study.

The FS System detects biological particles independently of the molecular biological techniques or the specific biochemistry involved. Since the capture is based on the physicochemical properties of the biological particle and is not based on the detection of any specific chemical compounds that build it, Fluid-Screen is ideal for the capture of unknown biological material, not normally identified by molecular biological techniques. This enables the search for alien biology as well as the search for new species of microbial life here on Earth (such as viruses that use 2,6-diaminopurine as a genetic base substitute for adenine [49–52]), and even for the possibility of the detection of the hypothesized "Shadow Biosphere" [53].

*4.2. Miniaturization and the Overall Engineering Adaptation for Venus and Other Space Applications*

While in its current form the Fluid-Screen System is exclusively developed for pharmaceutical and biotech industries and it is not yet space ready, the necessary miniaturization of the FS System and other required engineering adaptations are feasible.

The instrument would need to be optimized for small size and low weight. Such adaptations are feasible. The FS System is also a power-consuming instrument in its high-power configuration (Table 2); therefore, the simultaneous development of particle capture and separation protocols that rely on a lower peak power budget and lower flow rates would need to be developed.

To adapt to operation in a moving object, such as a balloon-based VAIHL mission, the FS System and the interfaced microscope, would need to be focused and permanently fixed before launch to avoid movement. They would also need to automatically focus during the mission. Interfacing with other instruments, including an externally developed primary reservoir, microscope or mass spectrometer would need to be developed separately and are currently being evaluated.

The FS System would need to be connected to the sample collector that collects liquid droplets from the atmosphere. Such sample collector technology is currently not available but could in principle be derived from cloud droplet capturing technologies known for Earth, e.g., "fog harp" instruments (e.g., [38–41]) that collect the sample into the primary reservoir. The inlet tube would connect the chip to the primary reservoir. The challenges of the sample collection are not only engineering in nature. Collectors that capture samples in bulk, and as a result, mix the contents of individual droplets, risk changing the individual chemical environment of each droplet, and therefore, prevent any studies of the heterogeneity of the droplets. If there are any cell-like structures, lipid vesicles or solid biological particles suspended in the liquid sulfuric acid droplets that could in principle be captured with the "Flow Chip" the mixing of the sample might destroy them or change their properties to a degree that captures would not happen. Capture methods that preserve the chemical microenvironment of the individual particles are currently not available and should be investigated.

Ideally, the FS System could connect to the LDMS-type instrument and present the sample directly for analysis. Such a configuration is challenging, as the LDMS has a high vacuum chamber, with vacuum levels at $5 \times 10^{-8}$ mbar, for sample analysis, introducing a major design challenge for any sample delivery system [37].

## 5. Conclusions

Further testing is required to establish suitable protocols for particle capture in environments mimicking Venusian cloud conditions. Specifically, concentrated sulfuric acid or solutions with high salt concentrations are a serious challenge for dielectrophoretic particle capture and separation and, if possible, will require the modification of the standard FS protocols.

While Venus' clouds are a particularly challenging environment for DEP-based biological particle capture methods, other bodies of the solar system with liquid water present, such as Jupiter's moon Europa or Saturn's moon Enceladus might be especially suitable for Fluid-Screen. DEP-based direct capture, separation and concentration of biological material are applicable to any environment where liquid water is present and should be considered for any space mission with astrobiological science objectives. Fluid-Screen System is a highly versatile instrument with proven capabilities. Upon further laboratory testing and modification, it will be ready to investigate and potentially resolve many of the outstanding astrobiological questions. While the current configuration of the FS System is not yet space ready, the necessary miniaturization of the FS System and adaptations are feasible.

**Author Contributions:** Conceptualization, M.U.W., J.J.P., R.E.W.; writing—original draft preparation, M.U.W., J.J.P., R.E.W.; writing—review and editing, M.U.W., J.J.P., R.E.W. All authors have read and agreed to the published version of the manuscript.

**Funding:** This research was funded by Fluid-Screen and the Massachusetts Institute of Technology.

**Institutional Review Board Statement:** Not applicable.

**Informed Consent Statement:** Not applicable.

**Data Availability Statement:** Not applicable.

**Acknowledgments:** We thank the extended Venus Life Finder Mission team (https://venuscloudlife.com/ , accessed on 1 November 2022) for useful discussions. We also thank Fluid-Screen and Hener teams.

**Conflicts of Interest:** The authors have been employed by Fluid-Screen (M.U.W., R.E.W.) and/or Hener, and have future cash and/or stock payments coming from their continued employment with Fluid-Screen and or Hener (M.U.W., R.E.W.). The authors also hold patents on the Fluid-Screen technology (M.U.W., R.E.W.). J.J.P. is an owner of JJ Scientific, a company that collaborates with Hener.

## References

1. Pohl, H.A.; Hawk, I. Separation of living and dead cells by dielectrophoresis. *Science* **1966**, *152*, 647–649.
2. Sher, L.D. Dielectrophoresis in lossy dielectric media. *Nature* **1968**, *220*, 695–696.
3. Voldman, J. Electrical forces for microscale cell manipulation. *Annu. Rev. Biomed. Eng.* **2006**, *8*, 425–454. https://doi.org/10.1146/annurev.bioeng.8.061505.095739.
4. Fernandez, R.E.; Rohani, A.; Farmehini, V.; Swami, N.S. Microbial analysis in dielectrophoretic microfluidic systems. *Anal. Chim. Acta* **2017**, *966*, 11–33.
5. Zhang, H.; Chang, H.; Neuzil, P. DEP-on-a-chip: Dielectrophoresis applied to microfluidic platforms. *Micromachines* **2019**, *10*, 423.
6. Sarno, B.; Heineck, D.; Heller, M.J.; Ibsen, S. Dielectrophoresis: Developments and applications from 2010 to 2020. *Electrophoresis* **2021**, *42*, 539–564. https://doi.org/10.1002/elps.202000156.
7. Cheng, I.-F.; Chen, T.-Y.; Lu, R.-J.; Wu, H.-W. Rapid identification of bacteria utilizing amplified dielectrophoretic force-assisted nanoparticle-induced surface-enhanced Raman spectroscopy. *Nanoscale Res. Lett.* **2014**, *9*, 324.
8. Schröder, U.-C.; Ramoji, A.; Glaser, U.; Sachse, S.; Leiterer, C.; Csaki, A.; Hübner, U.; Fritzsche, W.; Pfister, W.; Bauer, M. Combined dielectrophoresis—Raman setup for the classification of pathogens recovered from the urinary tract. *Anal. Chem.* **2013**, *85*, 10717–10724.
9. Yang, J.; Huang, Y.; Wang, X.-B.; Becker, F.F.; Gascoyne, P.R.C. Differential analysis of human leukocytes by dielectrophoretic field-flow-fractionation. *Biophys. J.* **2000**, *78*, 2680–2689.
10. Chang, S.; Cho, Y.-H. A continuous size-dependent particle separator using a negative dielectrophoretic virtual pillar array. *Lab Chip* **2008**, *8*, 1930–1936.
11. Huang, Y.; Holzel, R.; Pethig, R.; Wang, X.-B. Differences in the AC electrodynamics of viable and non-viable yeast cells determined through combined dielectrophoresis and electrorotation studies. *Phys. Med. Biol.* **1992**, *37*, 1499.
12. Abd Rahman, N.; Ibrahim, F.; Yafouz, B. Dielectrophoresis for biomedical sciences applications: A review. *Sensors* **2017**, *17*, 449.
13. Kwizera, E.A.; Sun, M.; White, A.M.; Li, J.; He, X. Methods of generating dielectrophoretic force for microfluidic manipulation of bioparticles. *ACS Biomater. Sci. Eng.* **2021**, *7*, 2043–2063.
14. Mohd Maidin, N.N.; Buyong, M.R.; Rahim, A.R.; Mohamed, M.A. Dielectrophoresis applications in biomedical field and future perspectives in biomedical technology. *Electrophoresis* **2021**, *42*, 2033–2059.
15. Weber, M.U.; Petkowski, J.J.; Weber, R.E.; Krajnik, B.; Stemplewski, S.; Panek, M.; Dziubak, T.; Mrozińska, P.; Piela, A.; Lo, S.L.; et al. Fluid-Screen Dielectrophoretic Microbial Capture, Separation and Detection I: Theoretical Study. *Nanotechnology* 2022, *submitted*.
16. Weber, M.U.; Petkowski, J.J.; Weber, R.E.; Krajnik, B.; Stemplewski, S.; Panek, M.; Dziubak, T.; Mrozińska, P.; Piela, A.; Reed, M.A. Fluid-Screen Dielectrophoretic Microbial Capture, Separation and Detection II: Experimental Study. *Nanotechnology* 2022, *submitted*.
17. Weber, R.E.; Petkowski, J.J.; Michaels, B.; Wisniewski, K.; Piela, A.; Antoszczyk, S.; Weber, M.U. Fluid-Screen as a Real Time Dielectrophoretic Method for Universal Microbial Capture. *Sci. Rep.* **2021**, *11*, 22222. https://doi.org/10.1038/s41598-021-01600-z.
18. Kwak, T.J.; Jung, H.; Allen, B.D.; Demirel, M.C.; Chang, W.-J. Dielectrophoretic separation of randomly shaped protein particles. *Sep. Purif. Technol.* **2021**, *262*, 118280.
19. Camacho-Alanis, F.; Ros, A. Protein dielectrophoresis and the link to dielectric properties. *Bioanalysis* **2015**, *7*, 353–371.
20. Mohamad, A.S.; Hamzah, R.; Hoettges, K.F.; Hughes, M.P. A dielectrophoresis-impedance method for protein detection and analysis. *AIP Adv.* **2017**, *7*, 15202.
21. Sonnenberg, A.; Marciniak, J.Y.; Krishnan, R.; Heller, M.J. Dielectrophoretic isolation of DNA and nanoparticles from blood. *Electrophoresis* **2012**, *33*, 2482–2490.
22. Song, Y.; Sonnenberg, A.; Heaney, Y.; Heller, M.J. Device for dielectrophoretic separation and collection of nanoparticles and DNA under high conductance conditions. *Electrophoresis* **2015**, *36*, 1107–1114.
23. Seager, S.; Petkowski, J.J.; Carr, C.E.; Grinspoon, D.; Ehlmann, B.; Saikia, S.J.; Agrawal, R.; Buchanan, W.; Weber, M.U.; French, R. Venus Life Finder Mission Study. *arXiv* **2021**, arXiv:2112.05153.
24. Seager, S.; Petkowski, J.J.; Carr, C.E.; Grinspoon, D.H.; Ehlmann, B.L.; Saikia, S.J.; Agrawal, R.; Buchanan, W.P.; Weber, M.U.; French, R.; et al. Venus Life Finder Missions Motivation and Summary. *Aerospace* **2022**, *9*, 385. https://doi.org/10.3390/aerospace9070385.
25. Agrawal, R.; Buchanan, W.P.; Arora, A.; Girija, A.P.; de Jong, M.; Seager, S.; Petkowski, J.J.; Saikia, S.J.; Carr, C.E.; Grinspoon, D.H.; et al. Mission Architecture to Characterize Habitability of Venus Cloud Layers via an Aerial Platform. *Aerospace* **2022**, *9*, 359. https://doi.org/10.3390/aerospace9070359.
26. Buchanan, W.P.; de Jong, M.; Agrawal, R.; Petkowski, J.J.; Arora, A.; Saikia, S.J.; Seager, S.; Longuski, J. Aerial Platform Design Options for a Life-Finding Mission at Venus. *Aerospace* **2022**, *9*, 363. https://doi.org/10.3390/aerospace9070363.
27. Seager, S.; Petkowski, J.J.; Carr, C.E.; Saikia, S.J.; Agrawal, R.; Buchanan, W.P.; Grinspoon, D.H.; Weber, M.U.; Klupar, P.; Worden, S.P.; et al. Venus Life Finder Habitability Mission: Motivation, Science Objectives, and Instrumentation. *Aerospace* 2022, *in review*.
28. Knollenberg, R.G.; Hunten, D.M. Clouds of Venus: A preliminary assessment of microstructure. *Science* **1979**, *205*, 70–74.
29. Knollenberg, R.G. A reexamination of the evidence for large, solid particles in the clouds of Venus. *Icarus* **1984**, *57*, 161–183.
30. Knollenberg, R.G.; Hunten, D.M. The microphysics of the clouds of Venus: Results of the Pioneer Venus particle size spectrometer experiment. *J. Geophys. Res. Space Phys.* **1980**, *85*, 8039–8058.
31. Knollenberg, R.G.; Hunten, D.M. Clouds of Venus: Particle size distribution measurements. *Science* **1979**, *203*, 792–795.

32. Knollenberg, R.G. Clouds and hazes. *Nature* **1982**, *296*, 18.
33. Knollenberg, R.; Travis, L.; Tomasko, M.; Smith, P.; Ragent, B.; Esposito, L.; McCleese, D.; Martonchik, J.; Beer, R. The clouds of Venus: A synthesis report. *J. Geophys. Res. Space Phys.* **1980**, *85*, 8059–8081.
34. Bains, W.; Petkowski, J.J.; Rimmer, P.B.; Seager, S. Production of Ammonia Makes Venusian Clouds Habitable and Explains Observed Cloud-Level Chemical Anomalies. *Proc. Natl. Acad. Sci. USA* **2021**, *118*, e2110889118.
35. Beegle, L.; Bhartia, R.; White, M.; DeFlores, L.; Abbey, W.; Wu, Y.-H.; Cameron, B.; Moore, J.; Fries, M.; Burton, A. SHERLOC: Scanning habitable environments with Raman & luminescence for organics & chemicals. In Proceedings of the 2015 IEEE Aerospace Conference, Big Sky, MT, USA, 7–14 March 2015; pp. 1–11.
36. Bhartia, R.; Beegle, L.W.; DeFlores, L.; Abbey, W.; Hollis, J.R.; Uckert, K.; Monacelli, B.; Edgett, K.S.; Kennedy, M.R.; Sylvia, M. Perseverance's Scanning Habitable Environments with Raman and Luminescence for Organics and Chemicals (SHERLOC) investigation. *Space Sci. Rev.* **2021**, *217*, 58.
37. Ligterink, N.F.W.; Kipfer, K.A.; Gruchola, S.; Boeren, N.J.; Keresztes Schmidt, P.; de Koning, C.P.; Tulej, M.; Wurz, P.; Riedo, A. The ORIGIN Space Instrument for Detecting Biosignatures and Habitability Indicators on a Venus Life Finder Mission. *Aerospace* **2022**, *9*, 312.
38. Knapczyk-Korczak, J.; Szewczyk, P.K.; Ura, D.P.; Bailey, R.J.; Bilotti, E.; Stachewicz, U. Improving water harvesting efficiency of fog collectors with electrospun random and aligned Polyvinylidene fluoride (PVDF) fibers. *Sustain. Mater. Technol.* **2020**, *25*, e00191.
39. Fernandez, D.M.; Torregrosa, A.; Weiss-Penzias, P.S.; Zhang, B.J.; Sorensen, D.; Cohen, R.E.; McKinley, G.H.; Kleingartner, J.; Oliphant, A.; Bowman, M. Fog water collection effectiveness: Mesh intercomparisons. *Aerosol Air Qual. Res.* **2018**, *18*, 270–283.
40. Schemenauer, R.S.; Joe, P.I. The collection efficiency of a massive fog collector. *Atmos. Res.* **1989**, *24*, 53–69.
41. Shi, W.; van der Sloot, T.W.; Hart, B.J.; Kennedy, B.S.; Boreyko, J.B. Harps Enable Water Harvesting under Light Fog Conditions. *Adv. Sustain. Syst.* **2020**, *4*, 2000040.
42. Baumgardner, D.; Fisher, T.; Newton, R.; Roden, C.; Zmarzly, P.; Seager, S.; Petkowski, J.J.; Carr, C.E.; Špaček, J.; Benner, S.A.; et al. Deducing the Composition of Venus Cloud Particles with the Autofluorescence Nephelometer (AFN). *Aerospace* **2022**, *9*, 492.
43. French, R.; Mandy, C.; Hunter, R.; Mosleh, E.; Sinclair, D.; Beck, P.; Seager, S.; Petkowski, J.J.; Carr, C.E.; Grinspoon, D.H.; et al. Rocket Lab Mission to Venus. *Aerospace* **2022**, *9*, 445. https://doi.org/10.3390/aerospace9080445.
44. Weber, M. AC Kinetics System for Universal Bacterial Capture. Ph.D. Thesis, Yale University, New Haven, CT, USA, 2017.
45. McMahon, S.; Cosmidis, J. False biosignatures on Mars: Anticipating ambiguity. *J. Geol. Soc. Lond.* **2022**, *179*. https://doi.org/10.1144/jgs2021-050.
46. Schulze-Makuch, D.; Grinspoon, D.H.; Abbas, O.; Irwin, L.N.; Bullock, M.A. A sulfur-based survival strategy for putative phototrophic life in the Venusian atmosphere. *Astrobiology* **2004**, *4*, 11–18.
47. Darling, H.E. Conductivity of sulfuric acid solutions. *J. Chem. Eng. Data* **1964**, *9*, 421–426.
48. Weast, R.C. (Ed.). *CRC Handbook of Chemistry and Physics*; CRC Press: Boca Raton, FL, USA, 1989; Volume 70, ISBN 0849304857.
49. Zhou, Y.; Xu, X.; Wei, Y.; Cheng, Y.; Guo, Y.; Khudyakov, I.; Liu, F.; He, P.; Song, Z.; Li, Z. A widespread pathway for substitution of adenine by diaminopurine in phage genomes. *Science* **2021**, *372*, 512–516.
50. Kirnos, M.D.; Khudyakov, I.Y.; Alexandrushkina, N.I.; Vanyushin, B.F. 2-Aminoadenine is an adenine substituting for a base in S-2L cyanophage DNA. *Nature* **1977**, *270*, 369–370.
51. Pezo, V.; Jaziri, F.; Bourguignon, P.-Y.; Louis, D.; Jacobs-Sera, D.; Rozenski, J.; Pochet, S.; Herdewijn, P.; Hatfull, G.F.; Kaminski, P.-A. Noncanonical DNA polymerization by aminoadenine-based siphoviruses. *Science* **2021**, *372*, 520–524.
52. Sleiman, D.; Garcia, P.S.; Lagune, M.; Loc'h, J.; Haouz, A.; Taib, N.; Röthlisberger, P.; Gribaldo, S.; Marlière, P.; Kaminski, P.A. A third purine biosynthetic pathway encoded by aminoadenine-based viral DNA genomes. *Science* **2021**, *372*, 516–520.
53. Davies, P.C.W.; Benner, S.A.; Cleland, C.E.; Lineweaver, C.H.; McKay, C.P.; Wolfe-Simon, F. Signatures of a Shadow Biosphere. *Astrobiology* **2009**, *9*, 241–249. https://doi.org/10.1089/ast.2008.0251.